\documentclass[letterpaper]{article} 
\usepackage{aaai2026}  
\usepackage{times}  
\usepackage{helvet}  
\usepackage{courier}  
\usepackage[hyphens]{url}  
\usepackage{graphicx} 
\usepackage{natbib}  
\usepackage{caption} 
\usepackage{algorithm}
\usepackage{algorithmic}

\usepackage{newfloat}
\usepackage{listings}
\DeclareCaptionStyle{ruled}{labelfont=normalfont,labelsep=colon,strut=off} 
\floatstyle{ruled}
\newfloat{listing}{tb}{lst}{}
\floatname{listing}{Listing}
\author {
    Jianqing Zhang\textsuperscript{\rm 1}\thanks{Work done during the internship at Tencent},
    Wei Xia\textsuperscript{\rm 2}\equalcontrib,
    Hande Dong\textsuperscript{\rm 2},
    Qiang Lin\textsuperscript{\rm 2},
    Jian Cao\textsuperscript{\rm 1}\equalcontrib
}
\affiliations {
    \textsuperscript{\rm 1}Shanghai Jiao Tong University \\
    \textsuperscript{\rm 2}Tencent \\
    tsingz@sjtu.edu.cn, xwellxia@tencent.com, cao-jian@sjtu.edu.cn
}

\usepackage{bibentry}

\usepackage{xspace}
\def\apo{\texttt{AP2O}\xspace}
\def\apocoder{\texttt{AP2O-Coder}\xspace}

\makeatletter
\DeclareRobustCommand\onedot{\futurelet\@let@token\@onedot}
\def\@onedot{\ifx\@let@token.\else.\null\fi\xspace}
\def\eg{\emph{e.g}\onedot} 
\def\ie{\emph{i.e}\onedot}

\def\wrt{\emph{w.r.t}\onedot} 

\makeatother

\definecolor{blue_}{RGB}{76, 114, 176}
\definecolor{orange_}{RGB}{221, 132, 82}
\definecolor{upload}{RGB}{47, 85, 151}
\definecolor{download}{RGB}{241, 13, 208}
\definecolor{red_}{RGB}{255, 0, 0}
\definecolor{gray_}{RGB}{127, 127, 127}
\definecolor{green_}{RGB}{1, 128, 0}
\definecolor{sjtured_}{RGB}{192, 0, 0}
\definecolor{sjtugreen_}{RGB}{84, 130, 53}
\definecolor{hist_red}{RGB}{194, 82, 83}
\definecolor{hist_blue}{RGB}{83, 110, 174}

\newcommand{\argmin}{\mathop{\arg\min}}
\newcommand{\linecode}[1]{\colorbox[rgb]{1,1,1}{\color{black} \texttt{#1}}}

\usepackage{amsthm,amsmath,amssymb}
\usepackage{cleveref}

\crefname{section}{Sec.}{Secs.}
\crefname{table}{Table}{Tables}
\crefname{figure}{Figure}{Figures}
\crefname{equation}{Eq.}{Eqs.}

\usepackage{booktabs}
\usepackage{xcolor}

\usepackage{pifont}
\usepackage{makecell}
\usepackage{colortbl}
\definecolor{grayline}{gray}{0.9}

\usepackage{array}
\usepackage{multirow}
\usepackage{subfigure}

\title{AP2O-Coder: Adaptively Progressive Preference Optimization for Reducing Compilation and Runtime Errors in LLM-Generated Code}

\begin{document}

\maketitle

\begin{abstract}
LLMs' code generation capabilities have yielded substantial improvements in the effectiveness of programming tasks. However, LLM-generated code still suffers from compilation and runtime errors. Existing preference optimization methods primarily focus on enhancing LLMs' coding abilities using pass/fail signals in the preference data, overlooking the deep-level error types in the failed codes. 
To address this, we propose Adaptively Progressive Preference Optimization (\apo) for coding (\ie, \apocoder), a method that guides LLMs adaptively and methodically to reduce code errors for code generation. Specifically, we construct an error notebook from failed codes and progressively optimize the LLM to correct errors type by type. Furthermore, we adaptively replay error types to tailor to the LLM's evolving weaknesses throughout training. 
Through extensive experiments on both code and general LLMs (Llama, Qwen, and DeepSeek series) with parameters ranging from 0.5B to 34B, our \apocoder improves code generation performance by up to 3\% in $pass@k$ while using less preference data. 
\end{abstract}

\begin{links}
    \link{Code}{https://github.com/TsingZ0/AP2O}
\end{links}

\section{Introduction}

Among all the capabilities of large language models (LLMs), code generation is one of the most attractive abilities \cite{sheokand2025codemixbench, dou2024stepcoder}. However, LLM-generated code still suffers from compilation and runtime errors \cite{tambon2025bugs}, such as \linecode{SyntaxError} and \linecode{TypeError}. 
Reinforcement Learning with Verifiable Rewards (RLVR) is a powerful technique for post-training to correct pre-trained LLMs' weaknesses, particularly in the code domain \cite{yue2025does, zhao2025absolute, wang2025reinforcement}. 
It only requires the problem prompts and unit tests to construct training data, with no need for output answers (codes). The LLM can self-generate multiple answers for each problem and use the corresponding unit tests to verify the correctness of these answers, automatically obtaining pass/fail signals \cite{liurltf}. 

Nevertheless, online RL approaches are unstable during training due to the changing models or environments \cite{moskovitz2023confrontingrewardmodeloveroptimization}. 
As an offline method, Direct Preference Optimization (DPO) \cite{rafailov2023direct} was introduced as a more stable alternative that does not require reward models and can be easily applied with verifiable rewards. However, DPO and its variants \cite{liu2025survey, pattnaik2024enhancing, meng2024simpo, croitoru2025curriculum} with identical data utilization exhibit three key shortcomings in reducing self-generated code errors: (1) \textit{unawareness of code errors}, as preference data is constructed solely from pass/fail signals; (2) \textit{inability to focus on specific error types}, since errors appear randomly in each training batch; and (3) \textit{neglect of the LLM's changing weaknesses}, as DPO samples preference data only once, and the static training set is pre-constructed, failing to adapt to the LLM's updating ability during training process. 

To address these issues, we propose Adaptively Progressive Preference Optimization (\apo), which consists of \textit{progressive preference optimization} and \textit{adaptive error replay} modules. We integrate \apo into the code LLM training and sandbox evaluation pipeline, creating our \apocoder. Inspired by human error correction practices \cite{xu2023importance}, we treat the acquisition of only pass/fail signals in existing DPO-based methods as \textit{\textbf{taking exams}}, akin to grading exam papers. In DPO-based methods, LLMs are guided to reduce failed answers solely based on these pass/fail signals, making it difficult for the model to understand why, where, and how it fails. 
Therefore, our \apocoder first \textit{\textbf{analyzes the failed answers in the exam}} using a programming-language-specific analyzer (\eg, a Python interpreter), acting as an expert. After analyzing, we organize the errors into an error notebook, ordered by error frequency (ascending or descending). To mimic human correction practices and enhance code error correction effectiveness, we correct errors type by type based on this error notebook within the progressive preference optimization module. During the \textit{\textbf{correction}} process, as the LLM is updated at each training step, the previously collected error notebook may no longer fit its current weaknesses. To mitigate this, we introduce the adaptive error replay module, which periodically evaluates the LLM on a small validation set, akin to \textit{\textbf{taking small quizzes}}. This process identifies the error types in the LLM's current failed answers and replays these error types, enabling the LLM to better focus on and correct them.

Through the systematic process of \textit{exam}, \textit{analysis}, \textit{correction}, and \textit{quiz}, our \apocoder outperforms five state-of-the-art baselines by up to 3\% in $pass@k$ on EvalPlus \cite{liu2023your} and LiveCodeBench v6 \cite{jain2024livecodebench}. This improvement is achieved across code and general LLMs, including CodeLlama \cite{roziere2023code}, DeepSeek-Coder \cite{guo2024deepseek}, Qwen2.5-Coder \cite{hui2024qwen2}, Llama3 \cite{grattafiori2024llama}, Qwen2.5 \cite{team2024qwen2}, and Qwen3 \cite{yang2025qwen3}, with parameter sizes ranging from 0.5B to 34B. We also find that progressing from low-to-high (L2H) error frequency is better for small models (\eg, 0.5B), while high-to-low (H2L) progression is more effective for large models (\eg, 34B). Our \apocoder also requires a smaller amount of preference data, thanks to its organized and adaptive data utilization. 

Below, we summarize our contributions:

\begin{itemize}
\item We analyze existing offline preference optimization methods in reducing LLM-generated code errors and identify three shortcomings: (1) inability to focus on specific errors, (2) erratic error identification, and (3) neglect of the LLM's changing weaknesses.
\item We propose \apocoder, with \apo as its core, to address these shortcomings by devising progressive preference optimization and adaptive error replay modules with a systematic process of exam, analysis, correction, and quiz, mimicking human error correction practices. 
\item We evaluate \apocoder on EvalPlus and LiveCodeBench, using various LLM types and sizes ranging from 0.5B to 34B parameters, demonstrating up to 3\% improvement in $pass@k$ over baselines.
\end{itemize}

\section{Related Work}

\subsection{Post-Training for Code Generation}

LLM is becoming an essential tool and valuable companion for programming tasks, especially with the rise of code generation capabilities \cite{cursor2025, anthropicclaude2025}. There are three main post-training techniques for code generation tasks: (1) instruction tuning \cite{ma2024llamoco, weyssow2023exploring}, (2) model distillation \cite{chen2023personalized, sun2024enhancing}, and (3) reinforcement learning (RL) \cite{mu2024clarifygpt, gehring2025rlef}. 
Instruction tuning is a foundational approach for post-training tasks \cite{lai2025survey}, but it heavily depends on high-cost expert-written annotations, such as problem-code pairs for code generation \cite{wu2025inversecoder}. Although model distillation mitigates this by leveraging existing high-performance models, it suffers from issues like error propagation and data leakage \cite{lei2023comprehensive}. 
RL with human feedback (RLHF \cite{kirkunderstanding}) is another approach, though RLHF can be biased and conflicting \cite{xiao2024algorithmic, cheng2024reinforcement}. In contrast, RL with verifiable rewards (RLVR \cite{zhao2025absolute}) has garnered increasing attention in recent years. While online RL suffers from instability caused by model and environment shifts during training \cite{moskovitz2023confrontingrewardmodeloveroptimization}, offline methods—especially offline preference optimization—offer greater stability when managing self-generated code errors. Hence, this paper focuses on offline preference optimization. 

\subsection{Offline Preference Optimization}

There are a few offline preference optimization methods specifically proposed or evaluated for coding tasks \cite{liu2025survey, da2025agent}, with most research focused on mathematical tasks \cite{liu2025trust}. In this work, we review general offline preference optimization methods, particularly those related to data utilization. These can be categorized into two main approaches: (1) dynamic sampling \cite{rao2025dynamic, gee2025code} and (2) curriculum learning \cite{pattnaik2024enhancing, shi2025cudip, hou2025novo, li20252d}.
Dynamic sampling methods mainly focus on resampling or active learning. Most resampling approaches require complex data quality criteria \cite{hu2024most} or auxiliary models \cite{huang2025adaptive}, limiting their utility. Active learning approaches \cite{muldrew2024active, xia2025selection} typically require re-training after each data selection step or rely on external large oracle LLMs (\eg, GPT-4 \cite{achiam2023gpt}), resulting in high costs. On the other hand, curriculum learning requires easy/hard task criteria \cite{lin2025survey}, which are often absent and difficult to establish, particularly in complex post-training tasks like correcting self-generated code errors.

\section{Preliminaries}

\subsection{Problem Formulation}

For the code generation tasks, we are given problem prompts and unit tests. Then, preference data $\mathcal{P}=\{<x, y_w, y_l>^1, <x, y_w, y_l>^2, \ldots\}$ can be constructed by any LLM itself, where $x$ represents any problem prompt, $y_w$ is the preferred answer, and $y_l$ is the rejected answer \wrt the given $x$. 

Our objective is to design an offline preference optimization method $\mathcal{L}$ that optimizes a pre-trained LLM $\theta$ to correct self-generated compilation and runtime errors and enhance its code generation ability. Formally, our problem is defined as: $\theta^\star \leftarrow \argmin \mathcal{L}(\theta;\mathcal{P})$. 

\subsection{Direct Preference Optimization}

Among offline preference optimization methods, DPO \cite{rafailov2023direct} is both fundamental and widely used, so we begin with it. DPO does not require a critic model and a reward model. Instead, it directly leverages the contrastive relationships among preference pairs from $\mathcal{P}$. Based on the Bradley-Terry model \cite{bradley1952rank}, preference probability model that $y_w$ is preferred over $y_l$ is
\begin{equation}
    P(y_w \succ y_l|x)=\sigma \left( \beta\log \frac{\pi^* (y_w|x)}{\pi_{\rm ref} (y_w|x)} - \beta\log \frac{\pi^* (y_l|x)}{\pi_{\rm ref} (y_l|x)} \right), 
\end{equation}
where $\sigma(\cdot)$ is the sigmoid function, $\beta > 0$ is a hyperparameter, $\pi^*$ is the optimal policy, and $\pi_{\rm ref}$ is the reference policy. The sample-level DPO loss function used to optimize the policy $\pi_{\theta}$, parameterized by the LLM $\theta$, is defined as
\begin{equation}
\begin{aligned}
    \ell_{\rm DPO}(\theta; x, &y_w, y_l) = \\
    - \log \sigma &\left( \beta\log \frac{\pi_{\theta} (y_w|x)}{\pi_{\rm ref} (y_w|x)} - \beta\log \frac{\pi_{\theta} (y_l|x)}{\pi_{\rm ref} (y_l|x)} \right). \label{eq:l_dpo}
\end{aligned}
\end{equation}
The DPO objective over the entire preference dataset $\mathcal{P}$ is to minimize the following loss function: 
\begin{equation}
    \mathcal{L}_{\rm DPO}(\theta; \mathcal{P}) = \mathbb{E}_{(x, y_w, y_l) \sim \mathcal{P}} [\ell_{\rm DPO}(\theta; x, y_w, y_l)]. \label{eq:L_DPO}
\end{equation}

\section{Method}

\subsection{Motivation}

Upon further analysis of the sample-level (\cref{eq:l_dpo}) and dataset-level (\cref{eq:L_DPO}) DPO loss functions, we identify three key shortcomings for DPO in correcting LLM-generated code errors:
\begin{enumerate}
    \item \textbf{Unawareness of code errors.} Unit tests can easily identify passed and failed answers, forming chosen ($y_w$) and rejected ($y_l$) pairs. However, there is no clear criterion to assign proper (negative) rewards to different error types (\eg, \linecode{KeyError} vs. \linecode{TypeError}), and constructing chosen-rejected pairs specifically for code error correction becomes challenging. Moreover, it's difficult to assess which errors are easier or harder to correct, rendering curriculum DPO variants \textit{inapplicable}. 
    \item \textbf{Inability to focus on specific error types.} DPO constructs a static preference dataset $\mathcal{P}$ by randomly shuffling, optimizing \cref{eq:L_DPO} batch by batch. This leads the LLM to encounter unpredictable error types, causing confusion in code error correction.
    \item \textbf{Neglect of the LLM's changing weaknesses.} Optimizing over uniformly scattered preference pairs overlooks the LLM's changing weaknesses. This also leads to inefficient training, wasted effort on irrelevant samples, and, in the worst case, degradation of the LLM's existing capabilities.
\end{enumerate}

To address these shortcomings, inspired by human error correction practices \cite{yang2021testing}, we propose \apocoder for AI coding tasks to: (1) construct an error notebook by collecting and analyzing errors, (2) guide the LLM to focus on correcting errors type by type, and (3) adaptively adjust the focus through small quizzes to fit the LLM's current capacity. 

\subsection{Overview}

Our \apocoder enhances any LLM based on its initial personalized coding ability by having it generate code for problem prompts across $M$ problems, akin to taking exams; we then give pass/fail signals to the answers (codes) using unit tests. We further analyze these errors by counting the frequency of different errors to create an error notebook. Based on this error notebook, \apocoder improves the LLM by progressively guiding it to correct errors (via progressive preference optimization) and reinforcing running errors with small quizzes (via adaptive error replay). Specifically, \apocoder consists of four steps: (1) \textbf{\textit{code answer generation (exam)}}, (2) \textbf{\textit{error diagnosis (analysis)}}, (3) \textbf{\textit{progressive preference optimization (correction)}}, and (4) \textbf{\textit{adaptive error replay (quiz)}}. Our core \apo consists of two key steps: \textit{correction} and \textit{quiz}. 

\begin{figure}[t]
\centering
\includegraphics[width=\columnwidth]{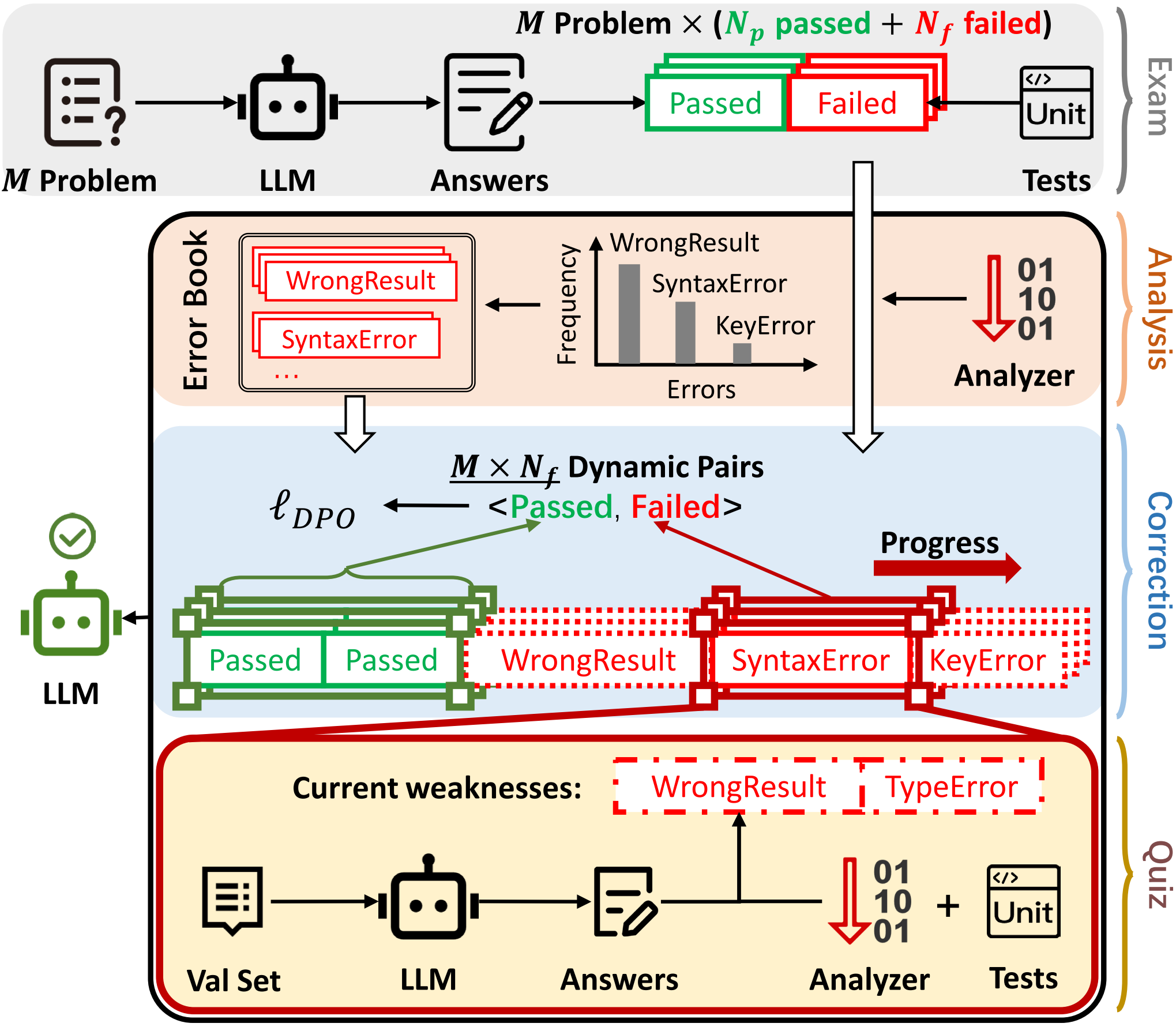}
\caption{The illustration of our \apocoder includes four steps: (1) code answer generation (exam), (2) error diagnosis (analysis), (3) progressive preference optimization (correction), and (4) adaptive error replay (quiz).}
\label{fig:apo}
\end{figure}

\subsection{Code Answer Generation (Exam)}

Initially, to assess a given LLM's baseline ability for subsequent targeted and personalized correction, we have the LLM take exams on $M$ coding problems $\{x^m\}_{m=1}^M$ and evaluate its answers using the corresponding multiple unit tests $\{(ut^1, ut^2, \ldots)^m\}_{m=1}^M$. Since it is difficult to gather a sufficient number of high-quality coding problems with unit tests, our \apocoder allow the LLM $\theta$ to generate $N$ answers (using a high temperature value\footnote{Following the widely used temperature setting for exploration \cite{shao2024deepseekmath}, our \apocoder set it to 1.0.}) for each problem to thoroughly explore the LLM's capability limits. Subsequently, our \apocoder obtains the grading results (pass or fail) for each problem, which serve as the \textit{intermediate} LLM-generated preference data, denoted as $\mathcal{D}_{\rm tr} = \{(x_, y^1_p, \ldots, y^{N_p}_p, y^1_f, \ldots, y^{N_f}_f)^m\}^M_{m=1}$, where $p$ and $f$ are short for passed and failed, respectively, and $N_p + N_f = N$. Formally, we have
\begin{equation}
    \mathcal{D}_{\rm tr} = \Gamma(\theta;\{x^m\}_{m=1}^M, \{(ut^1, ut^2, \ldots)^m\}_{m=1}^M), 
\end{equation}
where we use $\Gamma(\cdot)$ to represent the exam procedure. We illustrate this procedure by the \textit{Exam} part in \cref{fig:apo}. 

\subsection{Error Diagnosis (Analysis)}

Simply knowing whether an answer is correct or incorrect does not provide enough information for the LLM to improve itself, especially on complex tasks like code error correction. Inspired once again by human error correction practices \cite{xu2023importance}, we propose diagnosing failed answers through detailed error type analysis and organizing them into an \textit{\textbf{error notebook}}. However, the challenge lies in the need for an analyzer (expert) to perform the error diagnosis.

Fortunately, in domains like Python coding, interpreters can serve as experts, efficiently analyzing various errors with minimal effort. Specifically, we run the failed answers through a programming-language-specific analyzer, denoted as $\Psi(\cdot)$, to obtain detailed error type information. Formally, our \apocoder annotates the original $y_f^{m, n}$ with its corresponding \textit{ErrorType (E)} tag:
\begin{equation}
    y_{E}^{m, n} = \Psi(y_f^{m, n}), \forall n \in [N_f^m], m \in [M]. 
\end{equation}
Thus, we obtain a new error-notebook-structured $\mathcal{D}_{\rm tr}$, represented as
\begin{equation}
    \mathcal{D}_{\rm tr} = \{(x_, \{y^n_p\}_{n=1}^{N_p}, \{y_{E}^{ n}\}_{n=1}^{N_f})^m\}^M_{m=1}, 
\end{equation}
where the error frequency for each error type is also counted, as shown by the \textit{Analysis} part in \cref{fig:apo}. 
Note that high-frequency errors are not necessarily easy or hard to solve. A high-frequency error, \eg, a \linecode{SyntaxError}, may be easy for the LLM to correct, and once addressed, this error can be swiftly eliminated across massive problems and answers. 

\subsection{Progressive Preference Optimization (Correction)}

However, it remains challenging for the LLM to learn and correct errors from an unordered error notebook. Reflecting on human error correction practices, we humans typically \textit{prioritize error types} and \textit{correct them type by type}. Inspired by this, we propose sorting $y_f^{m, n}$ based on their error frequency in \apocoder. The sorting order—L2H or H2L—depends on the strength of the LLM's ability, where L2H indicates progression from low to high frequency, and H2L vice versa. 
Here, we consider \apocoder (H2L) as an example, as shown by the \textit{Correction} part in \cref{fig:apo}. 

In vanilla DPO and its variants, the training data is uniformly sampled, randomly shuffled, and static, resulting in three shortcomings, as discussed earlier. To address this, we propose a progressive preference optimization module that progressively focuses on correcting a specific type of error. 

Specifically, we construct an \textit{\textbf{error sliding window}} (with a width of $\lceil \frac{N_f^{m}}{T}\rceil$ and a depth of $M$, where $T$ is the total number of epochs) on the ordered list of failed answers across $M$ problems. For each problem $x$, we employ a dynamic-but-organized preference data construction approach to progressively select failed answers ($y_{E}$) with a specific type of error as the rejected samples. These are then paired with dynamically and randomly sampled passed answers ($y_p$) to form progressive preference data, denoted as $<x, y_p, y_{E}>$. Formally, we have
\begin{equation}
\begin{aligned}
    \mathcal{L}_{\rm AP2O-H2L}(\theta; \mathcal{D}_{\rm tr}) &= \\
    \mathbb{E}_{{E}\in \mathcal{E}} \mathbb{E}_{m \sim [M]} \mathbb{E}_{n \sim [N_p^m]} &\mathbb{E}_{n' \sim [N_{f,E}^m]}[\ell_{\rm DPO}(\theta; x^m, y_p^{m,n}, y_{E}^{m,n'})], \label{eq:apo1}
\end{aligned}
\end{equation}
where $\mathcal{E} = <E_1, E_2, \ldots>$ denotes the ordered error type list and $\varphi(E_1) > \varphi(E_2) > \cdots$. Here, $\varphi(\cdot)$ returns the frequency of a given error type $E$. We sample $E$ from $\mathcal{E}$ in order, and $N_{f,E}^m$ represents the size of the failed answer subset with error type $E$ for problem $m$. 

In the beginning, our \apocoder (H2L) focuses on correcting high-frequency errors, meaning the LLM encounters the same error across consecutive training steps, allowing it to concentrate on correcting \textit{a single type of error}. As training progresses, \apocoder (H2L) gradually shifts the error sliding window to focus on lower-frequency errors, exposing the LLM to a wider variety of errors in consecutive steps, thus \textit{enhancing generalization}. 

\begin{table*}[h]
\centering
\resizebox{!}{!}{
\begin{tabular}{l|ccc|ccc|cccccc}
\toprule
LLM Type & \multicolumn{3}{c|}{CodeLlama} & \multicolumn{3}{c|}{DeepSeek-Coder} & \multicolumn{6}{c}{Qwen2.5-Coder} \\
\midrule
LLM Size & 7B & 13B & 34B & 1.3B & 6.7B & 33B & 0.5B & 1.5B & 3B & 7B & 14B & 32B \\
\midrule
Init & 36.8 & 41.3 & 46.2 & 64.6 & 77.4 & 78.4 & 53.0 & 69.3 & 83.5 & 87.1 & 90.4 & 91.5\\
SFT-Coder & 37.9 & 43.2 & 46.8 & 64.8 & 75.9 & 78.9 & 60.1 & 70.4 & 85.1 & 87.4 & 90.7 &90.9\\
\midrule
DPO-Coder & 38.3 & 42.3 & 45.2 & 63.5 & 77.2 & 78.7 & 56.8 & 73.2 & 84.5 & 87.9 & 90.8 &91.0\\
Curri-DPO-Coder & 38.7 & 42.4 & 46.5 & 63.8 & 76.6 & 79.2 & 53.3 & 73.1 & 83.7 & 87.2 & 90.2 &90.8\\
Dyn-DPO-Coder & 38.6 & 42.3 & 44.9 & 63.4 & 76.2 & 78.8 & 57.1 & 71.5 & 84.7 & 87.6 & 90.7 &91.6\\
\rowcolor{grayline}
\apocoder (L2H) & \textbf{39.8} & 43.1 & 47.9 & \textbf{65.9} & 77.6 & 79.1 & \textbf{61.5} & \textbf{76.3} & 85.7 & 88.1 & 90.8 &91.8\\
\rowcolor{grayline}
\apocoder (H2L) & 38.9 & \textbf{44.5} & \textbf{49.6} & 64.7 & \textbf{78.8} & \textbf{80.1} & 56.5 & 71.7 & \textbf{86.3} & \textbf{88.9} & \textbf{91.4} &\textbf{92.2}\\
\bottomrule
\end{tabular}}
\caption{The $pass@1$ on EvalPlus (HumanEval) across various types and sizes of code LLMs.}
\label{tab:humaneval}
\end{table*}

\subsection{Adaptive Error Replay (Quiz)}

As the training process progresses, the LLM's ability changes. The current rule-abiding training data may no longer fit the LLM's changing weaknesses, leading to wasted effort on irrelevant samples and, at worst, potential degradation of its existing capabilities. 

To address this issue, we propose an adaptive error replay module to periodically evaluate the LLM's ability on a small validation set during the progressive preference optimization process, mimicking taking small quizzes. Originally, there is a validation dataset $\mathcal{D}_{\rm vl} = \{<x, y_p, y_f>^1, <x, y_p, y_f>^2, \ldots\}$ to evaluate a running model with unit tests and decide whether to save the current model as a checkpoint. Building on this existing training infrastructure, we apply the above analyzer to the answers generated on the validation set (one answer per validation problem), incurring negligible additional cost. Here, we do not calculate frequency but just get the ratio of each current error type. Then, we randomly sample $y_{E_{\rm vl}}$ from the entire failed answer list for each problem according to the ratio of the error type $E_{\rm vl}$. Subsequently, we replay these failed answers by adding them into the current error sliding window to give superiority to these failed answers, as they represent the current LLM's weaknesses. Formally, we update \cref{eq:apo1} to be 
\begin{equation}
\begin{aligned}
    \mathcal{L}_{\rm AP2O-H2L}(\theta; \mathcal{D}_{\rm tr}, &\mathcal{D}_{\rm vl}) = \\
    \mathbb{E}_{{E}\in \mathcal{E}} \mathbb{E}_{m \sim [M]} \mathbb{E}_{n \sim [N_p^m]}&\mathbb{E}_{n' \sim [N_{f,E}^m]} [\ell_{\rm DPO}(\theta; x^m, y_p^{m,n}, y_{E}^{m,n'}) + \\ &\ell_{\rm DPO}(\theta; x^m, y_p^{m,n}, y_{E_{\rm vl}}^{m,n'})], \\ 
\end{aligned}
\end{equation}
where $\{E_{\rm vl}^1, E_{\rm vl}^2, \ldots\} = \Phi(\theta, \mathcal{D}_{\rm vl})$ and $\Phi(\cdot)$ is the quiz procedure. 
We also guarantee that the number of total replayed failed answers is identical the width of the error sliding window to balance the current focusing and replayed data. We illustrate this procedure with the \textit{Quiz} part in \cref{fig:apo}. 

\section{Experiment}

\subsubsection{LLMs.} We evaluate the effectiveness of \apocoder by applying it to popular, state-of-the-art (SOTA) open-sourced code LLMs and general LLMs (Instruct versions) and post-training them to improve code generation performance. \textbf{Code LLMs}: CodeLlama \cite{roziere2023code}, DeepSeek-Coder \cite{guo2024deepseek}, and Qwen2.5-Coder \cite{hui2024qwen2}. \textbf{General LLMs}: Llama3 \cite{grattafiori2024llama}, Qwen2.5 \cite{team2024qwen2}, and Qwen3 \cite{yang2025qwen3}. We use LLMs ranging from 0.5B to 34B parameters.

\subsubsection{Baselines.} Since \apocoder operates as an offline preference optimization method that emphasizes progression through code preference data pairs, we select the following related baselines for comparison in the code domain. (1) \textit{Init}: The initial pre-trained (code) LLMs; (2) \textit{SFT-Coder}: Optimizing the pre-trained LLMs via supervised fine-tuning \cite{dodge2020fine} on coding tasks; (3) \textit{DPO-Coder}: Using DPO \cite{rafailov2023direct} with code-domain-specific training and sandbox evaluation pipelines; (4) \textit{Curri-DPO-Coder} \cite{pattnaik2024enhancing}: A representative curriculum DPO variant with code-specific pipelines; (5) \textit{Dyn-DPO-Coder} \cite{gee2025code}: A DPO variant that replaces the static preference dataset with dynamically sampled preference data during training progress. As for our \apocoder, we have two versions: \apocoder (L2H), and \apocoder (H2L), corresponding to two progression directions of the progressive preference optimization module. 

\subsubsection{Training Data.} Here, we focus on Python, one of the most frequently used programming languages. To obtain LLM-generated preference data, we use the coding problems and unit tests from the \textit{training/validation sets} of MBPP \cite{austin2021program} (384/90 problems for training/validation) and TACO \cite{li2023taco} (1678/420 problems for training/validation), respectively. We use MBPP by default. Since we focus on fine-grained learning from failed answers, we filter out coding problems with fewer than two failed answers. As the code answers are self-generated, the filter results are specific to the ability of the given LLMs but remains consistent across all baselines. 

\subsubsection{Other Settings.} Building on existing code LLM works \cite{team2024qwen2, hui2024qwen2}, we use popular benchmarks such as EvalPlus \cite{liu2023your} and LiveCodeBench v6 (Feb 2025–Apr 2025) \cite{jain2024livecodebench}, evaluating them with two metrics: $pass@k$ ($k\in\{1,5,10\}$) \cite{roziere2023code} and sample efficiency \cite{gao2022sample} with a temperature 0.6. Here, sample efficiency refers to the amount of data required during post-training. We conduct three training trials and report the average values. 
For more details and results, please refer to the Appendix. 

\subsection{Main Experiment}

\begin{table}[h]
\centering
\resizebox{\linewidth}{!}{
\begin{tabular}{l|ccc|ccc}
\toprule
Benchmark & \multicolumn{3}{c|}{MBPP} & \multicolumn{3}{c}{LiveCodeBench v6} \\
\midrule
LLM Size & 0.5B & 3B & 7B & 0.5B & 3B & 7B \\
\midrule
Init & 50.8 & 72.9 & 81.8 & 2.3 & 14.3 & 18.3\\
SFT-Coder & 55.4 & 74.5 & 82.4 & 2.9 & 14.7 & 18.2\\
\midrule
DPO-Coder & 51.9 & 76.0 & 83.5 & 2.9 & 14.8 & 18.4\\
Curri-DPO-Coder & 50.9 & 74.3 & 81.7 & 2.7 & 14.4 & 18.2\\
Dyn-DPO-Coder & 55.0 & 75.7 & 83.7 & 2.9 & 14.6 & 18.3\\
\rowcolor{grayline}
\apocoder (L2H) & \textbf{56.7} & \textbf{77.5} & 84.9 & \textbf{3.3} & 14.7 & 18.8\\
\rowcolor{grayline}
\apocoder (H2L) & 51.5 & 77.0 & \textbf{85.4} & 3.2 & \textbf{15.2} & \textbf{19.0}\\
\bottomrule
\end{tabular}}
\caption{The $pass@1$ on EvalPlus (MBPP) and LiveCodeBench across various sizes of Qwen2.5-Coder.}
\label{tab:mbpp_live}
\end{table}

Here, we select three widely used types of code LLMs along with their open-sourced versions of varying sizes. Due to space limitations, we present all types for the widely used EvalPlus (HumanEval) benchmark in \cref{tab:humaneval}, while showcasing only the strongest Qwen2.5-Coder for EvalPlus (MBPP) and LiveCodeBench v6 \cref{tab:mbpp_live}. 

\subsubsection{Pass Rate Comparison.} 

Across various types and sizes of code-specific LLMs, \apocoder consistently achieves superior pass rates, particularly for smaller models. It outperforms baselines by up to 3.1\% and improves over the pre-trained initial models by up to 8.5\% in \cref{tab:humaneval}. Even for well-pretrained large models (\eg, 30B+), \apocoder surpasses both baselines and initial models by up to 2.8\% and 3.4\%, respectively. 

In contrast to \apocoder's consistent improvements, all baselines occasionally degrade performance compared to the initial pre-trained models, particularly on large models. This is mainly due to the catastrophic forgetting problem \cite{li2025analyzing}, common in post-training methods (\eg, SFT and most DPO variants) \cite{fernando2024mitigating}, where continual optimization on a small post-training set leads to overfitting and loss of previously learned generalizable knowledge. Directly organizing the training data order, as done in Curri-DPO-Coder, also fails to mitigate this issue and may even exacerbate it, as shown in \cref{tab:mbpp_live}. 
Since \apocoder assesses the current capabilities of the gradually updating LLM and adaptively replays failed answers that fit its present weaknesses during the quiz phase, we can recover previously learned generalizable knowledge while effectively acquiring new knowledge. 

\subsubsection{Interesting Findings.} 

A deeper analysis of the pass rate reveals that \apocoder (L2H) outperforms \apocoder (H2L) on small and old models, whereas H2L performs better on larger and advanced models, as shown in \cref{tab:humaneval} and \cref{tab:mbpp_live}. This trend reflects the core distinction between “L” (low-frequency errors) and “H” (high-frequency errors): learning from diverse error types (L) enhances generalization, while repeated exposure to similar errors (H) promotes specialization. Thus, using L2H for weaker models—starting broadly then narrowing focus—risks convergence to local optima and poor generalization. For larger models, H2L and L2H perform similarly, but H2L yields better later-stage generalization by emphasizing low-frequency errors and exposing the model to more diverse error types in subsequent training. 
This distinction is both insightful and intriguing.

\subsection{Code Error Reduction}

\begin{figure}[h]
\centering
\includegraphics[width=\columnwidth]{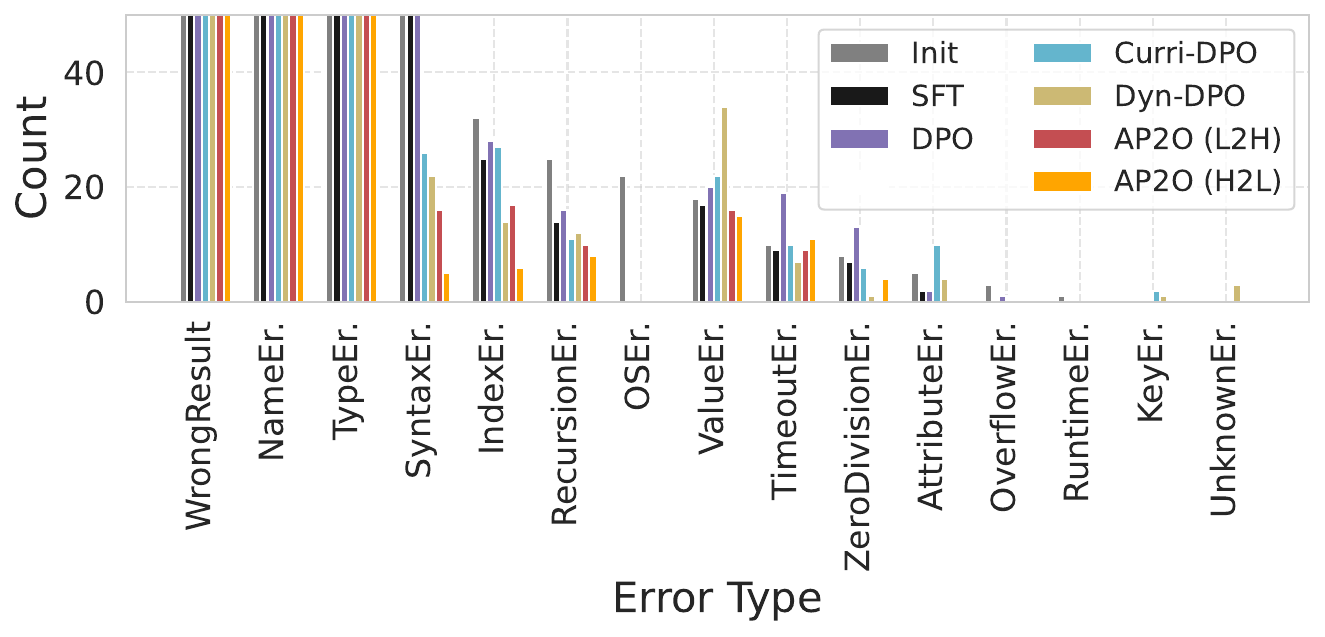}
\caption{The statistics of errors on the test benchmark using Qwen2.5-Coder-7B. ``Er.'' is short for ``Error''. }
\label{fig:error}
\end{figure}

\Cref{fig:error} provides interpretability into the effectiveness of our \apocoder. For better visibility, we clip the error count at 50. In general, the initial LLMs introduce the most errors, while post-training methods help reduce them, particularly for errors like \linecode{OSError}. However, these methods can also lead to regressions, such as increasing the frequency of certain error types like \linecode{ValueError}, or even introducing new types of errors, such as \linecode{KeyError}. In contrast, our \apocoder, with its progressive and adaptive modules, consistently reduces error counts without introducing new errors. Since the initial LLM here is the strong Qwen2.5-Coder-7B, \apocoder (H2L) performs better than \apocoder (L2H) by correcting a larger number of errors on high-frequency errors, although H2L may be slightly less impressive than L2H on low-frequency errors.

\subsubsection{Progressive Benefits.}

\begin{figure}[h]
\centering
\includegraphics[width=\columnwidth]{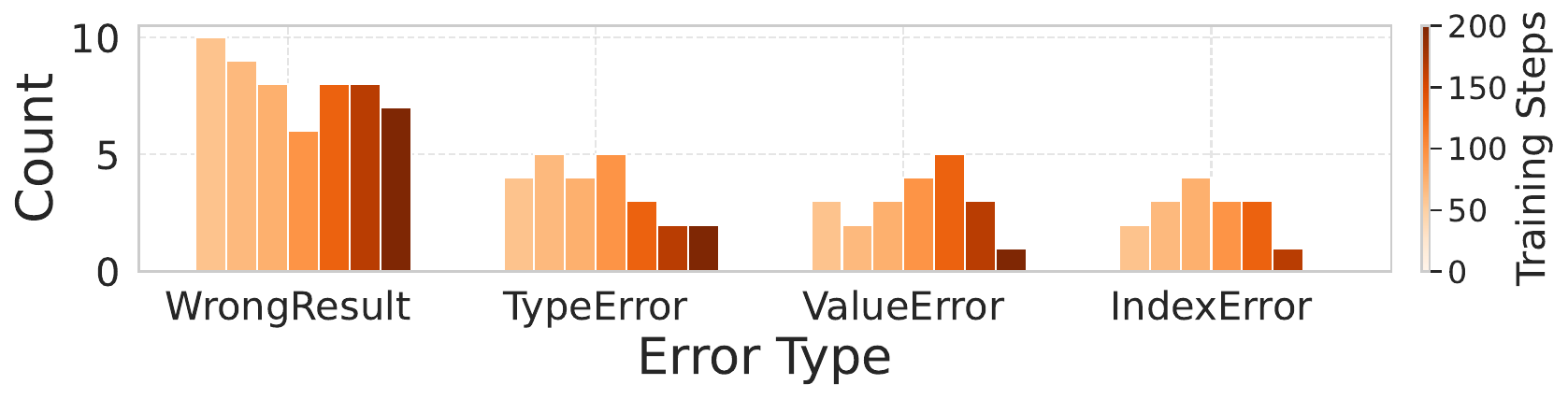}
\caption{The statistics of errors on the validation set during the quiz phase using Qwen2.5-Coder-7B. Our \apocoder progressively reduces errors.}
\label{fig:progress}
\end{figure}

To gain deeper insights into the training dynamics of \apocoder (H2L), we analyze the error reduction process in \cref{fig:progress}. Initially, there are a total of 19 failed answers dominated by the high-frequency \linecode{WrongResult} error in the validation set, along with several low-frequency error types. Following the H2L progressive strategy, \apocoder (H2L) first focuses on correcting the \linecode{WrongResult} errors while temporarily overlooking the others. As a result, the \linecode{WrongResult} is rapidly reduced in the early progression steps, but the less frequent errors may be negatively impacted. As training proceeds, \apocoder (H2L) shifts attention to the remaining low-frequency errors; however, this causes a resurgence in the \linecode{WrongResult} errors. Thanks to the adaptive error replay module, both high- and low-frequency errors are continually reduced, as the model revisits (prioritizes) the \linecode{WrongResult} errors while learning from the low-frequency ones. Notably, the \linecode{IndexError} is \textit{eliminated} by step 200 in \cref{fig:progress}.

\subsection{Generalization Ability on Large $k$ for $pass@k$.} 

\begin{figure}[h]
\centering
\includegraphics[width=\columnwidth]{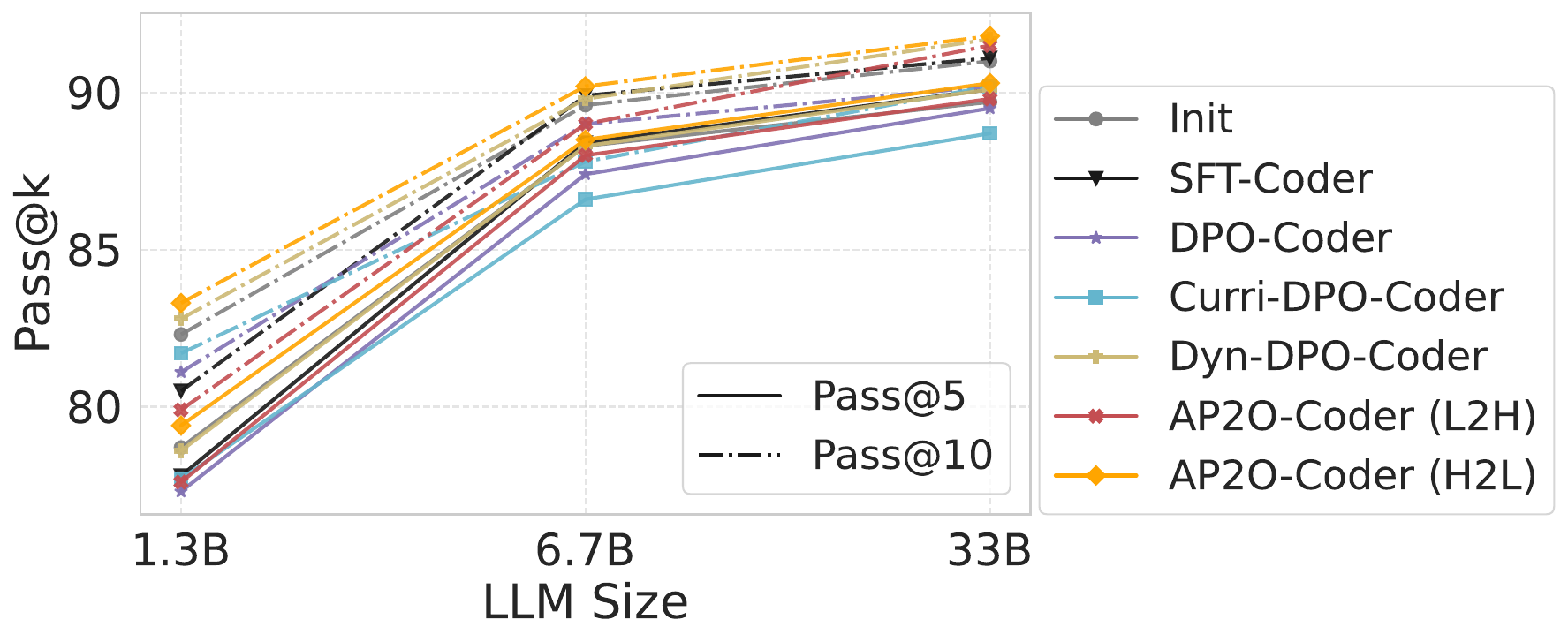}
\caption{The $pass@5$ and $pass@10$ on EvalPlus (HumanEval) using DeepSeek-Coder across various sizes.}
\label{fig:passk}
\end{figure}

We evaluate the generalization ability of post-trained code LLMs by benchmarking $pass@k$ for larger values of $k$ (\ie, $k \in {5, 10}$), as shown in \cref{fig:passk}. A commonly observed phenomenon in the literature is that post-training often improves $pass@1$ while degrading performance at higher $k$ values \cite{yue2025does, lyu2025top}. This is also evident in \cref{fig:passk}, where Curri-DPO-Coder shows significant performance degradation on large models, indicating that it may exacerbate the catastrophic forgetting problem. 
In contrast, our \apocoder (H2L) not only maintains improvements at $pass@1$ but also enhances generalization at larger $k$ values. This is attributed to its ability to emphasize low-frequency errors in the later stages of training.

\subsection{Sample Efficiency} 

\begin{figure}[h]
\centering
\includegraphics[width=\columnwidth]{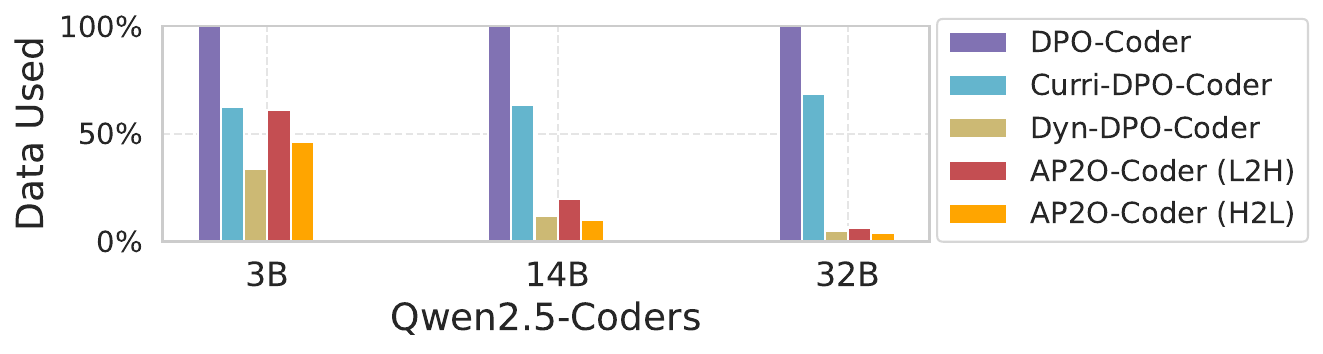}
\caption{The preference data pair requirements for training Qwen2.5-Coder across various sizes on the MBPP training set to achieve optimal performance.}
\label{fig:efficiency}
\end{figure}

In addition to improvements in pass rate, our \apocoder also demonstrates greater data efficiency, requiring only 4\%–60\% preference data pairs compared to the DPO's requirements, which is especially evident on large models. The H2L variant is more data-efficient than L2H, as prioritizing correcting high-frequency errors aligns with efficient human learning strategies \cite{larionova2022frequency}. As shown in \cref{fig:efficiency}, Curri-DPO-Coder exhibits the opposite trend, needing more data for larger models. Although Dyn-DPO-Coder uses the least data, its performance is poor, as shown in \cref{tab:humaneval} and \cref{tab:mbpp_live}.

\subsection{Adapting General LLMs to the Code Domain} 

\begin{figure}[h]
\centering
\includegraphics[width=\columnwidth]{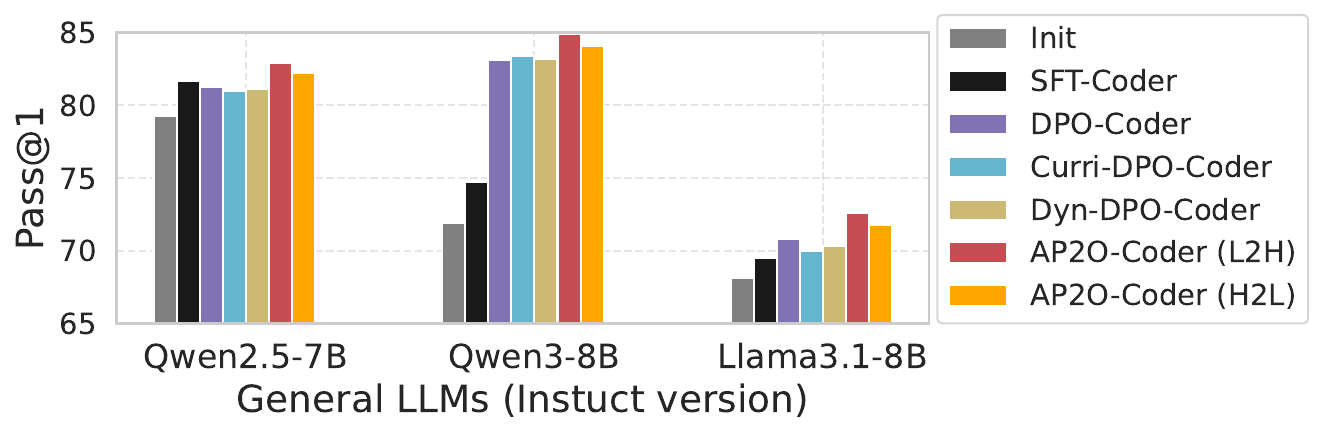}
\caption{The $pass@1$ on EvalPlus (MBPP) when adapting general LLMs, such as Qwen2.5, Qwen3, and Llama3, to the code domain. }
\label{fig:adapt}
\end{figure}

The evaluations above focus on existing code LLMs. Here, we demonstrate that our \apocoder can also effectively adapt pre-trained general LLMs to the code domain. Notably, some general models (\eg, Qwen3) are ``thinking'' models that tend to generate lengthy reasoning by default, leading to lower pass rates due to the 512-token budget constraint on the MBPP benchmark set by EvalPlus \cite{liu2023your}. While SFT struggles to mitigate this issue, the other offline preference optimization methods perform better in \cref{fig:adapt}. Among these, our \apocoder (L2H) achieves the best results, as the general 7B+ LLMs are poor in the specific code domain and benefit more from low-to-high-frequency (L2H) progressive optimization.

\subsection{Another Training Set} 

\begin{figure}[h]
\centering
\includegraphics[width=\columnwidth]{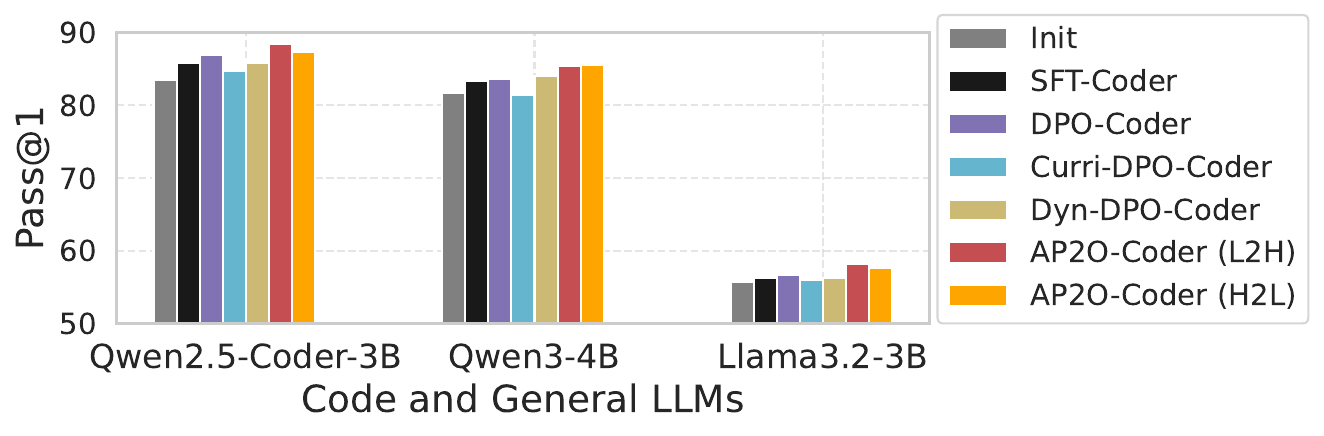}
\caption{The $pass@1$ on EvalPlus (HumanEval) after post-training on the TACO training set, using both code and general LLMs, such as Qwen2.5-Coder, Qwen3, and Llama3.}
\label{fig:another}
\end{figure}

We also demonstrate the robustness of our \apocoder by training on an alternative dataset, TACO, across both code and general LLMs. The performance trends of baselines remain consistent with previous results—Curri-DPO-Coder still yields relatively low pass rates. Notably, when using TACO, \apocoder (L2H) slightly outperforms \apocoder (H2L) on models with 3B+ parameters.

\subsection{Ablation Study} 

\begin{table}[h]
\centering
\resizebox{\linewidth}{!}{
\begin{tabular}{l|cc|cc}
\toprule
Benchmark & \multicolumn{2}{c|}{MBPP} & \multicolumn{2}{c}{LiveCodeBench} \\
\midrule
LLM Type & qw2.5-c & qw2.5 & qw2.5-c & qw2.5 \\
\midrule
\rowcolor{grayline}
\apocoder (L2H) & 84.9 & 82.9 & 18.8 & 15.1\\
$\quad -$ Adaptive Replay & 82.7 & 81.6 & 18.4 & 14.2\\
\rowcolor{grayline}
\apocoder (H2L) & 85.4 & 82.2 & 19.0 & 14.4\\
$\quad -$ Adaptive Replay & 82.1 & 81.0 & 18.5 & 13.6\\
\bottomrule
\end{tabular}}
\caption{The ablation study on EvalPlus (MBPP) and LiveCodeBench across various types of 7B+ code and general LLMs. We report the $pass@1$ results. ``qw2.5-c'' and ``qw2.5'' are abbreviations for Qwen2.5-Coder and Qwen2.5, respectively. }
\label{tab:ablation}
\end{table}

In the previous experiments, we have demonstrated the superiority of our \apocoder over the ablation variant Dyn-DPO-Coder. Here, we investigate additional ablation variants. Since the adaptive error replay module builds upon the progressive preference optimization module, we can only perform ablation by removing the adaptive replay component from both the L2H and H2L versions of \apocoder. As shown in \cref{tab:ablation}, this removal consistently leads to performance degradation across both code-specific and general LLMs, with performance in some cases falling below that of existing baselines (see \cref{tab:mbpp_live}). This confirms the role of the adaptive replay module in mitigating the catastrophic forgetting problem.

\subsection{Reward Curves}

\begin{figure}[h]
\centering
\includegraphics[width=\columnwidth]{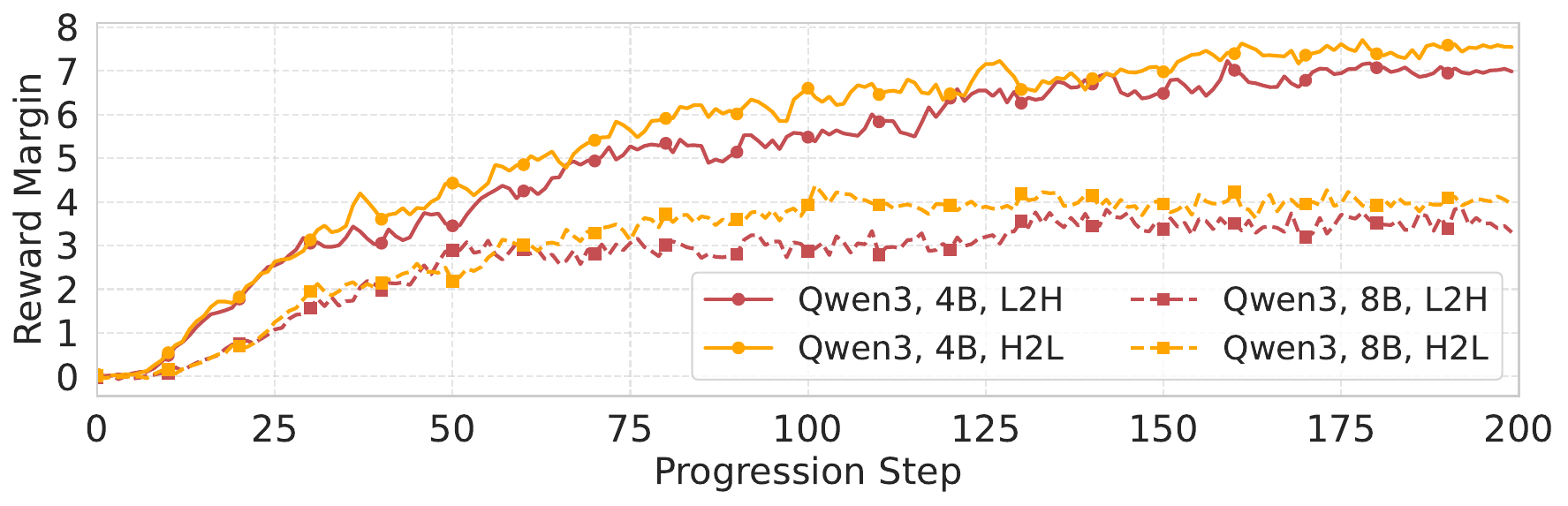}
\caption{Reward margins of chosen and rejected answers in \apocoder with Qwen3 during training.}
\label{fig:reward}
\end{figure}

Following prior work \cite{xiao2024cal}, we illustrate the training dynamics of our \apocoder in \cref{fig:reward}. The results show that optimization converges, with the H2L version outperforming L2H in reward measurement.


\section{Conclusion}

We propose an \apocoder that pioneers a human-inspired paradigm—exam, analysis, correction, quiz—to optimize LLMs to reduce LLM-generated code errors systematically. By introducing \apo that focuses on specific error types and continuously adapts the training data to the LLM's changing weaknesses, \apocoder achieves up to 3\% gains in $pass@k$ across diverse LLMs (0.5B–34B) while requiring less preference data. This advancement in error correction establishes a new state of the art, offering a robust, scalable solution to enhance code generation quality. 

\section{Acknowledgments}

This work is supported by the Shanghai Municipal Health Commission (grant number 2025ZHYL003).
The work is also sponsored by the CodeBuddy\footnote{https://www.codebuddy.ai} product. 

\bibliography{aaai2026}


\end{document}